\documentclass[prl, twocolumn,showpacs,nofootinbibfloatfix,amsmath,amsfonts,amssymb]{revtex4-1}%
\usepackage{amsmath,amsfonts,amssymb,color}
\usepackage{amsthm}
\usepackage{leftidx}
\usepackage{graphicx}
\usepackage{xcolor}
\usepackage{dcolumn}
\usepackage{bm}
\usepackage{epstopdf}
\usepackage{epsfig}
\usepackage{mathdots} 
\def\RB{\textcolor{black}}

\begin{document}
	\title{Square-root Floquet topological phases and time crystals}
	\author{Raditya Weda Bomantara}
	\email{Present Address: Physics Department, King Fahd University of Petroleum \& Minerals, Dhahran 31261, Saudi Arabia \newline \textbf{Email:} raditya.bomantara@kfupm.edu.sa}
	\affiliation{%
		Centre for Engineered Quantum Systems, School of Physics, University of Sydney, Sydney, New South Wales 2006, Australia
	}
	\date{\today}
	
	
	\vspace{2cm}
	
\begin{abstract}
Periodically driven (Floquet) phases are attractive due to their ability to host unique physical phenomena with no static counterparts. We propose a general approach in nontrivially devising a square-root version of existing Floquet phases, applicable both in noninteracting and interacting setting. The resulting systems are found to yield richer physics that is otherwise absent in the original counterparts and is robust against parameter imperfection. These include the emergence of Floquet topological superconductors with arbitrarily many zero, $\pi$, and $\pi/2$ edge modes, as well as $4T$-period Floquet time crystals in disordered and disorder-free systems ($T$ being the driving period). Remarkably, our approach can be repeated indefinitely to obtain a $2^n$th-root version of any periodically driven system, thus allowing for the discovery and systematic construction of exotic Floquet phases.  
	
\end{abstract}

\maketitle

\emph{Introduction.} It is recently proposed that by simply square-rooting an existing topological phase, a completely new material displaying exotic edge states properties is obtained \cite{sqrt1}. Inspired by Dirac's idea \cite{Dirac} in treating the Klein-Gordon equation \cite{Klein, Gordon}, such a square-rooting procedure is obtained by enlarging the degrees of freedom of the original system and devising a new Hamiltonian, the square of which yields two copies of the original system's Hamiltonian \cite{sqrt1}. In the last few years, various proposals of square-root topological phases have been theoretically made \cite{sqrt3,sqrt4,sqrt5,sqrt6,sqrt7,sqrt8,sqrt11,sqrt12,sqrt13,sqrt14} and experimentally verified \cite{sqrt2,sqrt9,sqrt10}. These studies however concern only the physics of static systems. 

Since the last decade, various phases of matter that can only be found in periodically driven systems (hereafter referred to as Floquet systems) have been identified and gained significant attention \cite{Flor3,Flor4,Flor6,Flor7,Flor11,Flor12,Flor15,Flor18,Flor19,Flor23,Flor24,Hoi,FMF6,FMF7}. Apart from being of fundamental interest, novel Floquet phases have been demonstrated to yield advantages in quantum information processing \cite{FMF2,FMF3,FMF4,RWB5,Floqc1,Floqc2,Floqc3}. It is thus envisioned that the possibility of square-rooting these Floquet systems will lead to even more exotic phases of matter with a significant quantum technological impact. To the best of our knowledge, however, such square-root Floquet phases have not been explored up to this date.

A static (Floquet) system is characterized by a Hermitian Hamiltonian $H$ (unitary one-period evolution operator $U$). This fundamental difference renders any known technique in square-rooting static systems inapplicable for use in Floquet setting. At first glance, square-rooting a Floquet system might even appear trivial. Indeed, by writing $U=e^{-\mathrm{i} H_{\rm eff} T}$ for some effective Hamiltonian $H_{\rm eff}$, its square-root is obtained simply through $H_{\rm eff}\rightarrow \frac{H_{\rm eff}}{2}$. However, it is important to note that $H_{\rm eff}$ is generically not physically accessible, especially for Floquet phases that have no static counterpart \cite{Flor8,Rud,Zhu,RWB,FMF1,FMF5}. In this case, simply reducing all system parameters by a half is not equivalent to $H_{\rm eff}\rightarrow \frac{H_{\rm eff}}{2}$ and will thus not yield the desired square-rooted system.

In this paper, we propose a general procedure for square-rooting a Floquet system in a systematic way, allowing its repetition to further generate any $2^n$th-root version of the system. Remarkably, unlike existing square-rooting procedures that typically only work for specific single-particle static topological systems, our proposal is applicable both to noninteracting and interacting Floquet systems, as demonstrated in the two explicit systems studied below. These case studies further reveal that such a square-rooting procedure is especially fruitful to yield systems with exotic properties that are otherwise absent in their original counterparts. For these reasons, our proposal opens an exciting opportunity to discover and study a variety of new Floquet phases.

\begin{center}
    \begin{figure}
        \includegraphics[scale=0.5]{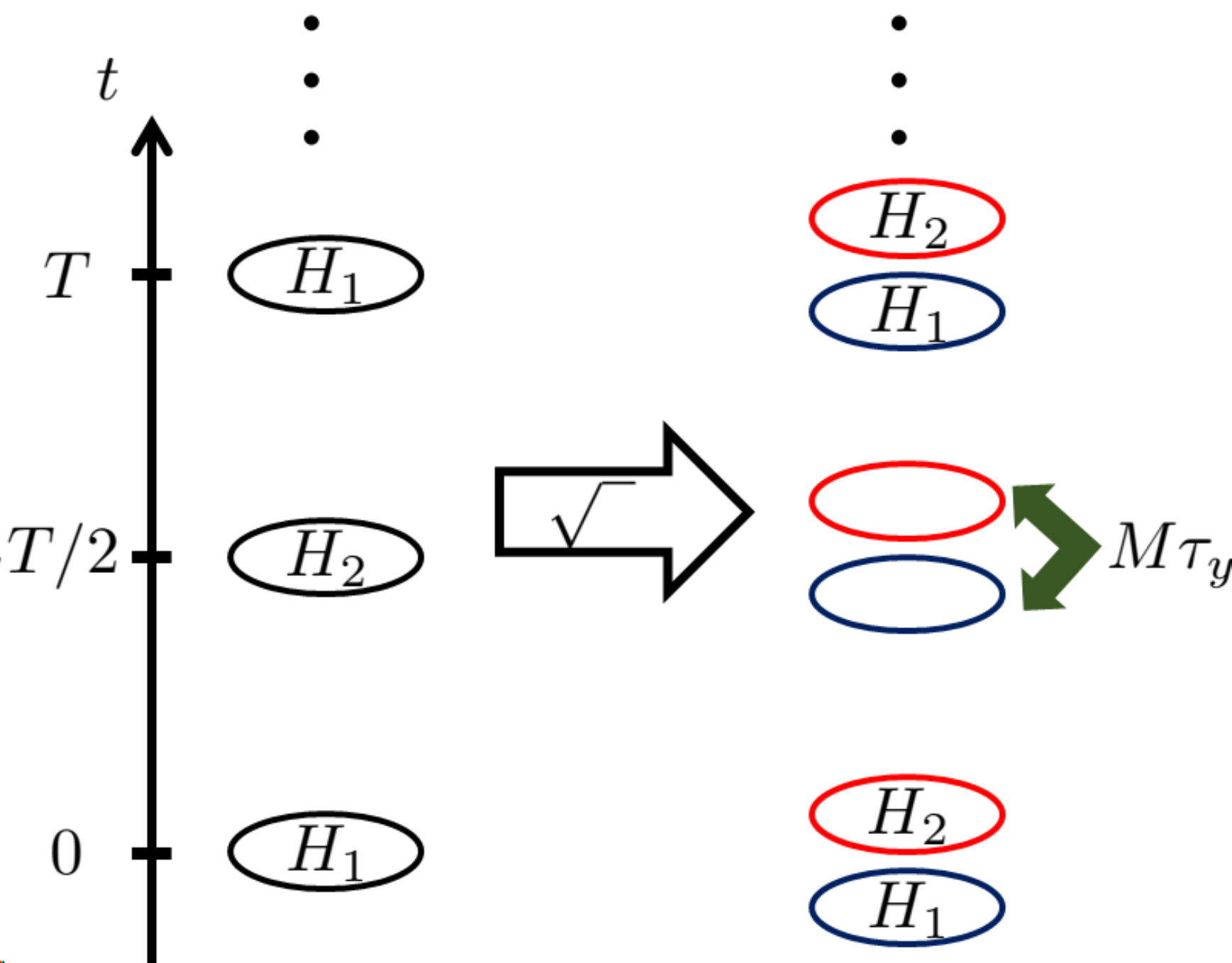}
        \caption{Construction of a square-root Floquet system from any time-periodic parent Hamiltonian, where $H_1\equiv H(t=0)$ and $H_2\equiv H(t=\frac{T}{2})$.}
        \label{fig:0}
    \end{figure}
\end{center}

\emph{General construction.} Note that the one-period evolution operator \RB{(hereafter referred to as Floquet operator)} of \emph{any} Floquet system can be written as
\begin{equation}
    U = U_2 U_1 =\left(\mathcal{T}e^{-\mathrm{i} \int_{0}^{T/2} H(t+T/2) dt} \right) \left(\mathcal{T}e^{-\mathrm{i} \int_{0}^{T/2} H(t) dt} \right) \;, \label{cond}
\end{equation}
where $H(t)=H(t+T)$ is the system's Hamiltonian of period $T$ and $\mathcal{T}$ the time-ordering operator. To obtain its square-root version, we define a two-time-step Hamiltonian 
\begin{equation}
    h_{(1/2)}(t) =\begin{cases}
    \begin{array}{c}
    H(t) \frac{1+\tau_z}{2} \\
    + H(t+\frac{T}{2}) \frac{1-\tau_z}{2}
    \end{array} & \text{ for } n<\frac{t}{T}\leq n+\frac{1}{2} \\
     M \tau_y & \text{ for } n+\frac{1}{2}<\frac{t}{T}\leq n+1
    \end{cases} \;, \label{method}
\end{equation}
where $M$ is a \RB{system independent} real parameter, $n\in \mathbb{Z}$, and $\tau_{x/y/z}$ are Pauli matrices representing an additional (pseudo)spin-1/2 degree of freedom. Such a square-root Hamiltonian and its original counterpart are schematically shown in Fig.~\ref{fig:0}. \RB{Our construction can be intuitively understood as follows. Consider a particle initially living in the subsystem $\tau_z=+1$. It evolves under $H(t)$ for the first half of the period and then moves to the other subsystem during the second half of the period. As the Hamiltonian repeats, the particle, which is now in the subsystem $\tau_z=-1$, evolves under $H(t+T/2)$ for another half period and moves back to the subsystem $\tau_z=+1$ at the end of the second period. That is, only after two periods will the particle experience the full Floquet operator of the parent system, while remaining on the same subsystem. On the other hand, the particle can exhibit a nontrivial evolution over a period to yield various new physics, some of which are highlighted in the case studies below.}

More explicitly, at $M T = \pi $, the Floquet operator associated with Eq.~(\ref{method}) takes the form
\begin{equation}
    u_{(1/2)}= \RB{e^{-\mathrm{i} \frac{\pi}{2} \tau_y} \left(\begin{array}{cc}
    	U_1 & \mathbf{0} \\
    	\mathbf{0} & U_2
    	\end{array} \right) } = \left(\begin{array}{cc}
         \mathbf{0} &  -U_2 \\
         U_1 & \mathbf{0}
    \end{array} \right) \;.
\end{equation}
In particular, $u_{(1/2)}^2=\mathrm{diag}(-U_2 U_1, -U_1 U_2)$, thereby reproducing two decoupled copies of the target $U$ (up to a unitary transformation). \RB{Its ability to host two different orderings of $U$ is particularly fruitful for potential parallel processing applications. For example, topological invariants of one-dimensional (1D) chiral symmetric Floquet topological systems are defined from the winding numbers of Floquet operators at two different orderings \cite{CFTI}. In such cases, the simultaneous realization of both Floquet operators with our construction can offer a significant speed-up in detecting their topological invariants.}

\RB{At $MT\neq \pi$, the diagonal elements of $u_{(1/2)}$ may in general become nonzero, while its off-diagonal elements are deformed away from $U_1$ and $U_2$. This renders $u_{(1/2)}^2$ no longer diagonal and directly related to $U$. However, by intentionally setting $MT\neq \pi$ in all the numerics below, we find that $u_{(1/2)}^2$ inherits the main physics of the target system. Therefore, Eq.~(\ref{method}) can still be regarded as the same square-root solution of the target model provided $MT-\pi$ is not too large. The insensitivity of our construction to the fine-tuning of $M$ demonstrates the robustness of square-root Floquet phases.}


Importantly, our square-rooting procedure is scalable, i.e., it can be applied indefinitely to obtain the $2^n$th-root of any Floquet system, thus opening avenues for obtaining a variety of Floquet phases with even more exotic physical properties. In the following, we explicitly apply our procedure on two representative systems. For conciseness, we only focus on their square-root counterparts, emphasizing the unique features not found in the corresponding parent systems. In Ref.~\cite{Supp}, the $4$th- and $8$th-root version of such systems are presented.


\emph{Square-root Floquet topological superconductor with arbitrarily many edge modes.} A remarkable feature of Floquet topological phases is their possibility to support any number of edge modes through appropriate choice of system parameters \cite{kk3,LW,LW2,LW3,LW4,RWB2,RWB3}. To demonstrate our square-rooting procedure at work, we consider the Floquet topological superconducting model introduced in Ref.~\cite{RWB2}, which is described by a two-time-step Hamiltonian switching between $H_1$ and $H_2$ at every half period \RB{(Two continuous variations of such a model are further considered in Ref.~\cite{Supp})}. There, 
\begin{equation}
    H_\ell = \sum_{j=1}^N \mu_\ell c_j^\dagger c_j +\sum_{j=1}^{N-1} \left(-J_\ell c_j^\dagger c_{j+1} +\Delta_\ell c_j^\dagger c_{j+1}^\dagger +h.c. \right)\;, \label{model1}
\end{equation}
where $\mu_\ell$, $J_\ell$, and $\Delta_\ell$ are respectively the chemical potentials, hopping amplitudes, and pairing strengths, $\ell=1,2$, $c_j$ is the fermionic operator at site $j$, and $N$ is the system size. In particular, at $\mu_2 = -2J_2 = -2\Delta_2 = m \mu_1= 2m J_1 = 2m\Delta_1$ with $m\in \mathbb{R}$, such a system supports $n$ pairs of Majorana zero modes and Majorana $\pi$ modes (MZMs and MPMs) for $n\pi<m<(n+\frac{1}{2})\pi$ \cite{RWB2}. Here, MZMs and MPMs are topologically protected Hermitian edge-localized operators which respectively commute and anticommute with the system's Floquet operator \cite{FMF2,FMF3,FMF4}. \RB{Even at the special parameter values above, such a system has a very complex and unphysical effective Hamiltonian of the form $H_{\rm eff}\propto \arccos\left[\cos(x\cos(k/2))\cos(m x\cos(k/2)) \right]$, where $x$ is a constant and $k$ the quasimomentum. Therefore, the trivial square-root procedure $H_{\rm eff}\rightarrow \frac{H_{\rm eff}}{2}$ is indeed infeasible.}

Following our general construction, the corresponding square-root system is obtained as a two-time-step Hamiltonian $h_{(1/2)}^{(FTSC)}(t)$ which switches between $h_{(1/2),1}^{(FTSC)}=h_1+h_2$ and $h_{(1/2),2}^{(FTSC)}=\sum_{j=1}^N M \mathrm{i} c_{1,j}^\dagger c_{2,j} +h.c.$ after every $T/2$ time interval, where $h_{\ell}$ with $\ell=1,2$ take the form of Eq.~(\ref{model1}) with $c_{j}\rightarrow c_{\ell,j}$. Physically, such a system represents a pair of $p$-wave superconductors with inter-chain hopping applied during the second half of the period. \RB{It can in principle be realized by proximitizing two chains of semiconducting wires with an $s$-wave superconductor, such that $h_1$ and $h_2$ are achieved through the same mechanism as that in the realization of Kitaev chain \cite{TSCexp1,TSCexp2}. The interchain hopping is further obtained and controlled by modulating the separation between the two chains.}

\begin{center}
    \begin{figure}
        \includegraphics[scale=0.5]{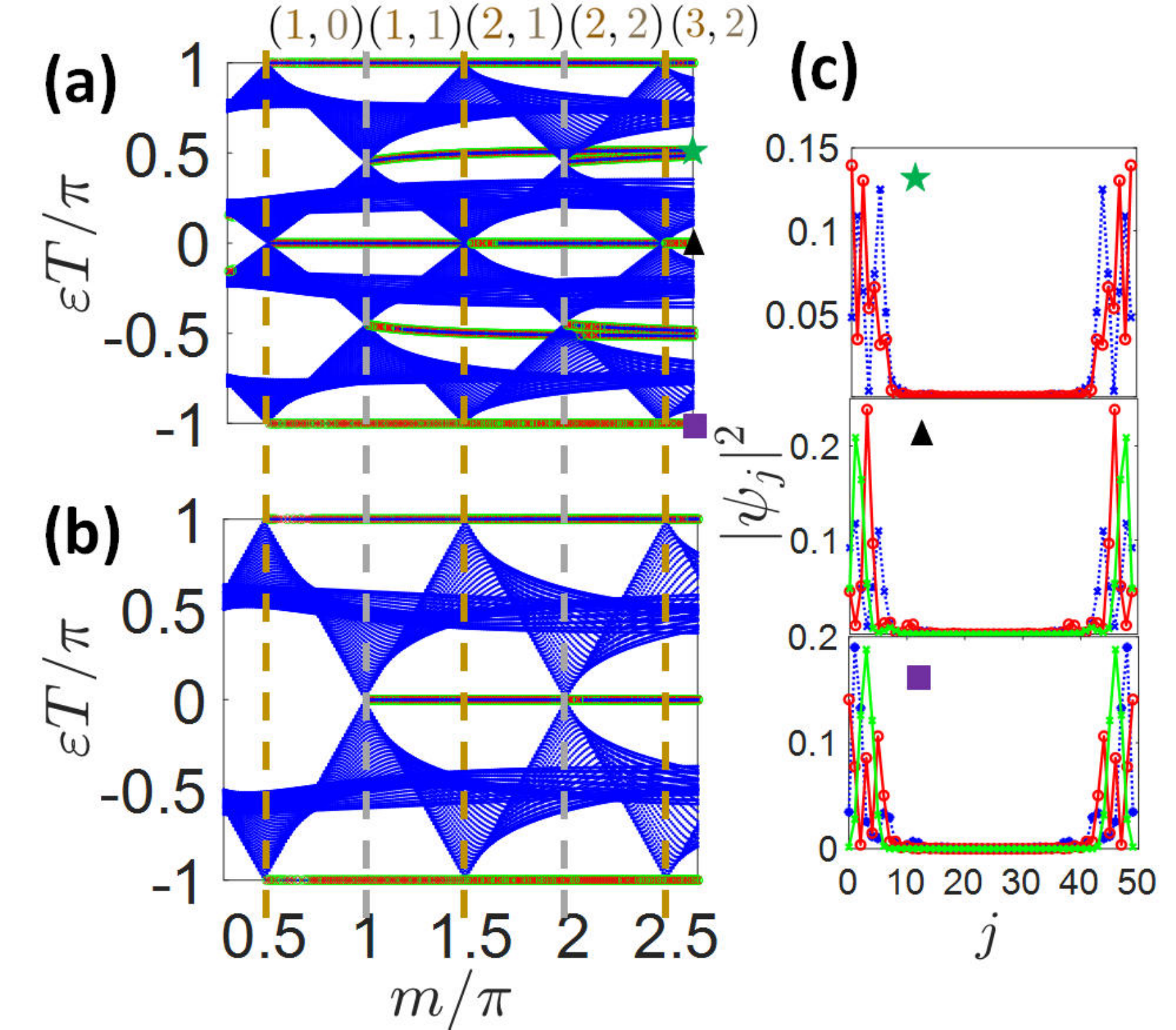}
        \caption{Quasienergy excitation spectrum of the (a) square-root and (b) original Floquet topological superconductor of Ref.~\cite{RWB3}. The vertical lines mark the topological phase transition points, while $(\nu_\pi,\nu_0)$ is a pair of topological invariants determining the number of MPMs and MZMs in the original system. (c) Typical wavefunction profiles of the system's edge modes. The system parameters are chosen as $\mu_2 T=m\mu_1 T= 2m$, $-J_2 T=mJ_1 T=1.05m$, $-\Delta_2 T=m\Delta_1 T = 0.95m$, $M T=0.9 \pi$, and $N=50$.}
        \label{fig:1}
    \end{figure}
\end{center}

The Bogoliubov–de Gennes (BdG) Floquet operator $\mathcal{U}$, which is related to the actual Floquet operator via $u=\frac{1}{2}\Psi^\dagger \mathcal{U} \Psi$, can be explicitly obtained \RB{and its exact form is detailed in Ref.~\cite{Supp}.} The system's quasienergy ($\varepsilon$) excitation spectrum is then obtained from the eigenvalues $e^{-\mathrm{i}\varepsilon T}$ of $\mathcal{U}$ and is summarized in Fig.~\ref{fig:1}(a). The associated quasienergy excitation spectrum of the original system is plotted in Fig.~\ref{fig:1}(b) for reference. Note in particular that both systems share the same topological phase transition points, marked by parameter values at which gap closing exists. MZMs and MPMs are associated with quasienergy zero and $\pi/T$ solutions respectively in Fig.~\ref{fig:1}(a,b). Moreover, the following two features are clearly observed.

First, the presence of MPMs in the original system leads to the simultaneous presence of MZMs and MPMs in the square-root system. This feature can further be analytically proven by computing a pair of topological invariants $(\nu_0,\nu_\pi)$ and $(\nu_0^{(1/2)},\nu_\pi^{(1/2)})$ for the original and square-root system respectively \cite{RWB2,Supp}. In particular, $\nu_0$ ($\nu_0^{(1/2)}$) and $\nu_\pi$ ($\nu_\pi^{(1/2)}$) respectively determine the number of pairs of MZMs and MPMs in the original (square-root) system. By leaving the technical detail in Ref.~\cite{Supp}, we indeed find that $\nu_\pi=\nu_0^{(1/2)}=\nu_\pi^{(1/2)}$, thus confirming our observation above.

Second, the presence of MZMs in the original system leads to the emergence of edge modes at $\approx \pi/(2T)$ quasienergy. Recently, it was shown that such $\pi/2$ modes may become parafermions \cite{par1,par2,par3} in the presence of interaction \cite{RWB4}. These $\pi/2$ modes are however not as ubiquitous as MZMs and MPMs, and their construction previously involves a rather elaborate driving scheme \cite{RWB4}. With our square-rooting procedure, such $\pi/2$ modes can be systematically generated and their origin traced back from the topology of the squared model. That is, while a topological invariant characterizing these $\pi/2$ modes in the square-root system directly is presently unknown to us, the presence of $\pi/2$ modes can still be inferred from the invariant $\nu_0$ defined on the squared system.


As elaborated in Ref.~\cite{Supp}, the presented square-root Floquet topological superconductor inherits the chiral symmetry of its parent system. This chiral symmetry is responsible for protecting MZMs and MPMs in the system. Indeed, even at imperfect square-root parameter $MT\neq \pi$, MZMs and MPMs remain pinned at $0$ and $\pi/T$ quasienergy respectively (see Fig.~\ref{fig:1}(a)). By contrast, the observed $\pi/2$ modes slightly deviate from the expected $\pi/(2T)$ quasienergy due to the absence of symmetry protection. It remains to be seen if the presence of interaction that promotes these $\pi/2$ modes to $Z_4$ parafermions in the ideal limit, such as via the mechanism elucidated in Ref.~\cite{RWB4}, may render them more robust against such imperfection effect. 

\emph{Square-root Floquet time crystals.} Our construction is not limited to single-particle systems. Indeed, it can be applied to square-root a Floquet time crystal (FTC), i.e., a many-body phase of matter characterized by robust subharmonic observable dynamics  \cite{DTC1,DTC2,DTC3,DTCexp1,DTCexp2,DTCexp3,DTCexp4,DTCexp5,DTCexp6,DTCexp7,DTCexp8,DTC4,DTC5,DTC6,DTC7,DTC8,DTC10,DTC11,DTC12,DTC13,DTC14,DTC15,Pizzi,DTC16,DTC17,DTC18,DTC19,DTC20,DTC21,DTCcm8,repDTC,DTCrel2,DTCnew}. Focusing first on the many-body localization (MBL) protected FTC model of Ref.~\cite{DTC2}, its square-root is obtained by plugging in
\begin{equation}
H(t)= h_j X_j \;, \;\;\;\; H(t+T/2)= J_j Z_j Z_{j+1}+h_j^Z Z_j   \label{example2}
\end{equation}
to Eq.~(\ref{method}). There, $P_j=J_j,h_j,M,$ and $h_{j}^Z$ are each randomly taken from a uniform set $[\bar{P}-\Delta P,\bar{P} +\Delta P]$,  $X_j$ and $Z_j$ are Pauli matrices associated with the $j$th site in the 1D lattice, and $\tau_{x/y/z}$ are additional Pauli matrices. As a rather unrealistic interpretation of the system, it describes a single spin ($\tau_{x/y/z}$) interacting with a 1D Ising model. In a more physical setting, it can be effectively and more robustly realized with two interacting and periodically driven spin-1/2 chains. As detailed in Ref.~\cite{Supp}, this is achieved by replacing $Z_j\rightarrow Z_{j,A}$, $X_j\rightarrow X_{j,A}X_{j,B}$, $\tau_z\rightarrow Z_{j,A}Z_{j,B}$, and $\tau_y\rightarrow \sum_{j=1}^N X_{j,B}$ in Eq.~(\ref{example2}), where $A,B$ label the two chains. The resulting system, which involves at most nearest-neighbor two-body interactions, can in turn be implemented with FTC successful trapped ions \cite{DTCexp1,DTCexp7} and superconducting circuit \cite{DTCexp8} platforms \cite{repDTC}.

\begin{center}
    \begin{figure}
        \includegraphics[scale=0.5]{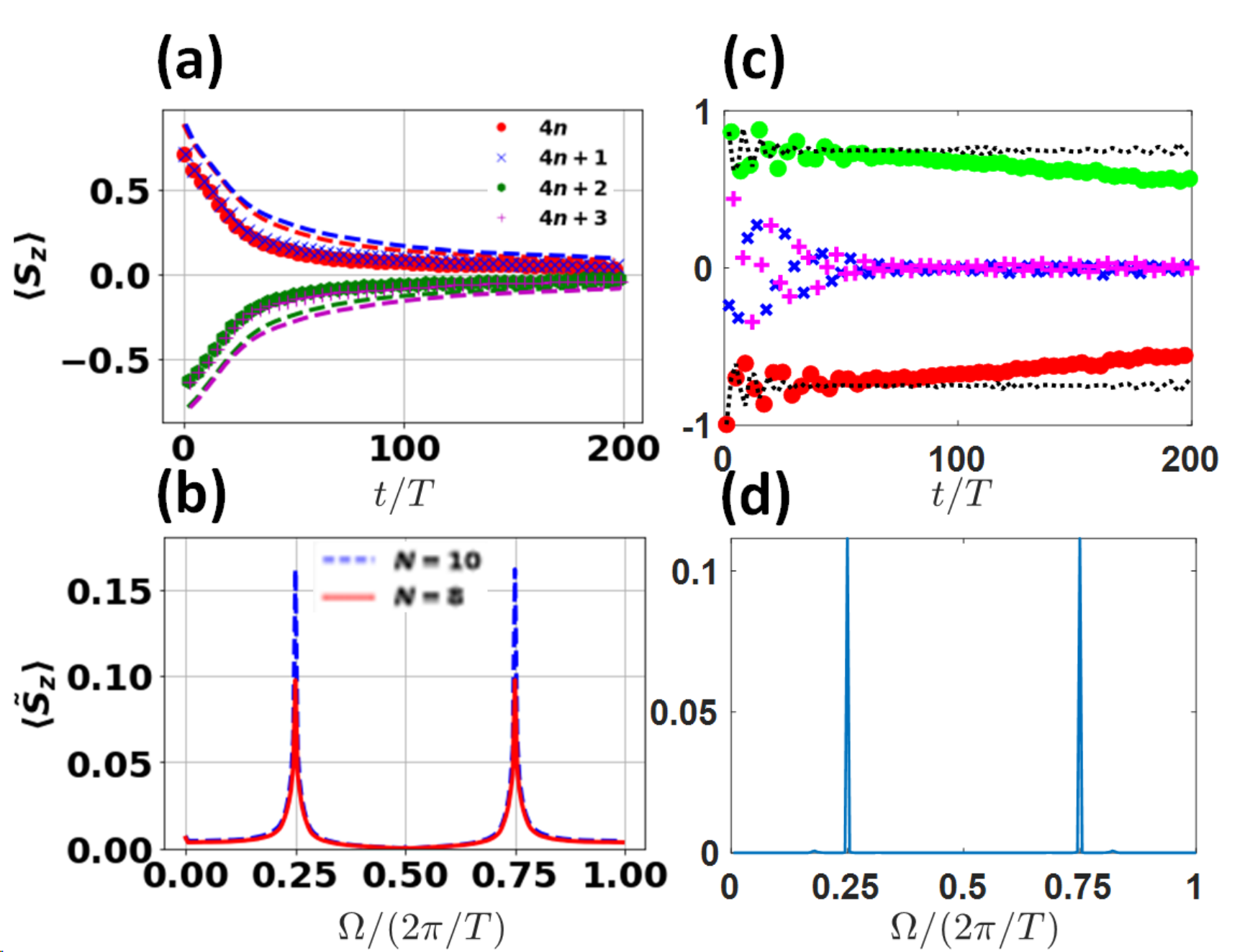}
        \caption{\RB{Stroboscopic magnetization profile (a,c) and its associated power spectrum (b,d) under square-root of (a,b) Eq.~(\ref{example2}) after averaging over $500$ disorder realizations for a system of $8$ (solid marks) and $10$ (dashed lines) particles, (c,d) Eq.~(\ref{example3}) for a system of $100$ particles. In panel (c), the black dotted lines show the corresponding profile under its parent system. The system parameters are chosen as (a,b) $\bar{h}T=0.92 \pi$, $\bar{J}T=2$, $\bar{h}^Z T = 0.3$, $\bar{M} T =0.95\pi$, $\Delta h T=0.05\pi$, $\Delta J T = 1$, $\Delta h^Z T =0.3$, and $\Delta M T=0.05\pi$, (c,d) $JT=1$, $hT=0.1$, $\phi T =\frac{0.9\pi}{2}$, and $M T = 0.98\pi$. Note that in all panels, the interaction strength $J$ is comparable with that used in previous studies of FTCs such as Refs.~\cite{DTC2,Pizzi}.}}
        \label{fig:2}
    \end{figure}
\end{center}

In Fig.~\ref{fig:2}(a,b), we plot the stroboscopic magnetization dynamics, i.e., $\langle S_z \rangle =\frac{1}{N}\sum_{j=1}^N \langle Z_j \rangle$, and its associated power spectrum, i.e.,  $\langle \tilde{S}_z\rangle = |\frac{1}{L} \sum_{m=\ell}^L \langle S_z\rangle e^{-\mathrm{\frac{\ell \Omega T}{L}}}|$, under the square-root of Eq.~(\ref{example2}) and a generic initial state $|\psi(0)\rangle = \prod_{j=1}^N e^{-\mathrm{i} \frac{\pi}{8} Y_j} |00\cdots 0\rangle$. \RB{Despite considerable deviation from the ideal values $MT=h_j T=\pi$, a robust $4T$- rather than $2T$-periodicity is observed, thus highlighting the system's nature as a square-root version of Ref.~\cite{DTC2}. Moreover, Fig.~\ref{fig:2}(b) reveals that this subharmonic behavior improves with the system size. These features imply that such a square-root model is indeed a genuine FTC.} 
It is also worth noting that this $4T$-period FTC is physically different from that proposed in Ref.~\cite{repDTC} due to the absence of $\Omega=\frac{\pi}{T}$ peak in Fig.~\ref{fig:2}(b) (cf. Fig.~2 of Ref.~\cite{repDTC}). 

The advantage of our square-rooting procedure over the FTC construction proposed in Ref.~\cite{repDTC} is the possibility to devise large-period FTCs without MBL. This can be accomplished, e.g., by square-rooting the disorder-free \RB{continuously driven} Lipkin-Meshkov-Glick (LMG) model. \RB{The latter is a variation of the kicked LMG model of Ref.~\cite{DTC8} in which the Dirac-delta driving is replaced by a harmonic driving. It is thus related to the model of Ref.~\cite{Pizzi}. Specifically, the parent Hamiltonian to be inserted in Eq.~(\ref{method}) takes the form 
\begin{equation}
H(t)=\left(\sum_{i,j} \frac{J}{2N} Z_i Z_j +\sum_i h X_i \right) \left(1+\cos(\omega t)\right) +\sum_i \phi X_i \;. \label{example3}
\end{equation}}


Since such a system preserves the total spin $\mathcal{S}^2=\sum_{i,j} \left(X_i X_j +Y_i Y_j +Z_i Z_j \right)$, numerical studies of very large system sizes are accessible via exact diagonalization. In Fig.~\ref{fig:2}(c,d), a clear and long-lived $4T$ oscillation profile is observed from $\langle S_z(t) \rangle$ and its power spectrum, taking $|\psi(0)\rangle =|00\cdots 0\rangle$ at $100$ particles as the initial state. By implementing the system via two interacting LMG chains as detailed in Ref.~\cite{Supp}, a long-lasting $4T$ oscillation profile is observable with merely $\sim 10$ particles, a much smaller system size than that required for observing a similar feature in Ref.~\cite{Pizzi} \RB{(see Ref.~\cite{Supp} for the underlying mechanism).} 

\RB{By repeating the square-root procedure to the above systems, disordered and clean $2^nT$-period FTCs can respectively be obtained (see, e.g., Ref.~\cite{Supp}). Importantly, they are also observable at system sizes accessible with current technology \cite{DTCexp1,DTCexp7,DTCexp8}, thus paving the way for experimentally exploring FTCs beyond their subharmonic signatures, e.g., confirmation of condensed matter phenomena in the time domain \cite{DTC21,DTCcm8}.}

\emph{Concluding remarks.} We have proposed a systematic and general construction of square root Floquet phases exhibiting exotic properties not found in the parent systems. We explicitly applied our procedure to obtain Floquet topological superconductors with arbitrarily many MZMs, MPMs, and the elusive $\pi/2$ modes, as well as FTCs beyond period-doubling. \RB{As a primary advantage of our construction, it amounts to coupling two copies of the parent systems and can thus be realized in the same platform as the latter, inheriting their feasibility. Indeed, our square-root FTCs can be directly implemented in the platforms of Ref.~\cite{DTCexp1,DTCexp7,DTCexp8} under the available resources \cite{Supp}. While the parent model of our square-root topological superconductor has not been experimentally realized, the predicted $\pi/2$ modes can arise from square-rooting a simpler model. In an upcoming work \cite{sqrtexp}, we will experimentally demonstrate the signature of $\pi/2$ modes in an acoustic square-root topological insulator.}

\RB{Our procedure can generate nontrivial square-root Floquet systems even if their parent systems have static counterparts. For example, a Hamiltonian of the form $H(t)=H_0(1+\sin(\omega t))$ has a simple Floquet operator $U=e^{-\mathrm{i} H_0 T}$ that can be trivially square-rooted through $H_0\rightarrow \frac{H_0}{2}$. However, by explicitly plugging in $H(t)$ to Eq.~(\ref{method}), the resulting Floquet operator instead yields a complex effective Hamiltonian that has no static counterpart \cite{Supp}.}

\RB{The above example demonstrates that a square-root Floquet system is not unique. Consequently, the proposed approach is not the only means of generating a square-root Floquet system, but it provides a motivation and inspiration for devising other square-rooting schemes.} In particular, an alternative procedure involving a smooth driving protocol rather than the two-time-step drive proposed here would be desirable. Another possible improvement of the current scheme is to exploit time degree of freedom to replace the ancillary Pauli matrices used in our construction. This may be achieved by adapting the technique proposed in Ref.~\cite{RWB}. Finally, in Ref.~\cite{nthroot}, we extend the present approach to generate an $n$th-root Floquet phase, where $n$ is any arbitrary integer.


\begin{acknowledgments}
	{\bf Acknowledgement}: This work is supported by the Australian Research Council Centre of Excellence for Engineered Quantum Systems (EQUS, CE170100009). The author thanks Longwen Zhou, Chinghua Lee, Jiangbin Gong, and Weiwei Zhu for helpful discussions. In particular, the author acknowledges Chinghua Lee for suggesting the potential parallel processing applications of the square-root procedure.
\end{acknowledgments}

\newpage 

\onecolumngrid 
\appendix

\begin{center}
    {\Large \textbf{Square-root Floquet topological phases and time crystals: Supplemental Material}}
\end{center}

This Supplemental Material contains five sections. In Sec.~A, we explicitly define and calculate $\nu_0^{(1/2)}$ and $\nu_\pi^{(1/2)}$ which respectively determine the number of pairs of MZMs and MPMs in the square-root Floquet topological superconductor considered in the main text. In Sec.~B, we apply the procedure introduced in the main text to obtain the square-root of two continuously driven topological superconducting systems. In Sec.~C, we present a more detailed analysis of our square-root FTCs which includes their physical and robust implementation by utilizing ideas from quantum error correction, origin of time crystallinity, and potential experimental realization. In Sec.~D, we demonstrate that our procedure yields a nontrivial square-root of $H(t)=H_0(1+\sin(\omega t))$. In Sec.~E, we construct and analyze the $4th$- and $8th$-root version of the systems considered in the main text.  

\section{A. Calculation of topological invariants $\nu_0^{(1/2)}$ and $\nu_\pi^{(1/2)}$}
\label{app}

\RB{We first note that the Hamiltonians describing the periodically driven topological superconductor in the main text can be written in the form (using $\left\lbrace c_{j,\ell}^\dagger , c_{j',\ell'}\right\rbrace= \delta_{j,j'}\delta_{\ell,\ell'}$, $\left\lbrace c_{j,\ell} , c_{j',\ell'}\right\rbrace= 0$, as well as ignoring constant terms)
\begin{eqnarray}
h_{(1/2),1}^{(FTSC)}&=& \sum_{\ell=1,2} \left[\sum_{j=1}^{N}  \frac{\mu_\ell}{2} \left(c_{j,\ell}^\dagger c_{j,\ell} - c_{j,\ell} c_{j,\ell}^\dagger \right) +\sum_{j=1}^{N-1} \left(\frac{J_\ell}{2} \left(c_{j,\ell}^\dagger c_{j+1,\ell} - c_{j+1,\ell} c_{j,\ell}^\dagger \right)+\frac{\Delta_\ell}{2} \left(c_{j,\ell}^\dagger c_{j+1,\ell}^\dagger - c_{j+1,\ell}^\dagger c_{j,\ell}^\dagger \right) +h.c. \right) \right] \;, \nonumber \\
h_{(1/2),2}^{(FTSC)}&=& \sum_{j=1}^N \frac{\mathrm{i} M}{2} \left(c_{j,1}^\dagger c_{j,2} - c_{j,2} c_{j,1}^\dagger \right) +h.c. \;.
\end{eqnarray}
during the first-half and second-half of the period respectively. For each of these Hamiltonians, we define the corresponding BdG Hamiltonian matrix $\mathcal{H}_{(1/2),s}^{(FTSC)}$ which satisfies $h_{(1/2),s}^{(FTSC)}=\frac{1}{2}\Psi^\dagger \mathcal{H}_{(1/2),s}^{(FTSC)} \Psi$, where $\Psi=\bigotimes_{j=1}^{N}\bigotimes_{\ell=1,2} \left(c_{j,\ell}, c_{j,\ell}^\dagger\right)^T$ is the Nambu operator. We further define $|j\rangle$ to be a vector selecting the $jth$ site component of $\Psi$, while $\sigma_{x/y/z}$ and $\tau_{x/y/z}$ respectively represent Pauli matrices acting on the particle-hole space and chain species directly. That is, 
\begin{eqnarray}
\Psi^\dagger \sigma_x |j\rangle \langle j'| \Psi &=& \sum_{\ell=1,2} \left(c_{j,\ell}^\dagger c_{j',\ell}^\dagger + c_{j,\ell} c_{j',\ell} \right) , \;\;\;\; \Psi^\dagger \sigma_{y} |j\rangle \langle j'| \Psi = \sum_{\ell=1,2} \mathrm{i} \left(c_{j,\ell}^\dagger c_{j',\ell}^\dagger - c_{j,\ell} c_{j',\ell} \right) , \nonumber \\ 
\Psi^\dagger \sigma_{z} |j\rangle \langle j' | \Psi &=& \sum_{\ell=1,2} \left(c_{j,\ell}^\dagger c_{j',\ell} - c_{j,\ell} c_{j',\ell}^\dagger \right) , \;\;\;\; \Psi^\dagger \tau_{x/y/z} |j\rangle \langle j' | \Psi =  \sum_{\ell,\ell'=1,2} \left(c_{j,\ell}^\dagger \left[\tau_{x/y/z}\right]_{\ell,\ell'} c_{j',\ell'} + c_{j,\ell} \left[\tau_{x/y/z}\right]_{\ell,\ell'} c_{j',\ell'}^\dagger \right) . \nonumber \\
\end{eqnarray}
We can then write the BdG Hamiltonians as
\begin{eqnarray}
\mathcal{H}_{(1/2),1}^{(FTSC)}&=& \sum_{\ell=1,2} \frac{1+(3-2\ell)\tau_z}{2} \left\lbrace\sum_{j=1}^{N}  \mu_\ell \sigma_z |j\rangle \langle j|  +\sum_{j=1}^{N-1} \left[\left(J_\ell \sigma_z -\mathrm{i} \Delta_\ell \sigma_y \right) |j\rangle \langle j+1 | +h.c. \right] \right\rbrace \;, \nonumber \\
\mathcal{H}_{(1/2),2}^{(FTSC)}&=& \sum_{j=1}^N M \tau_y |j\rangle \langle j| \;.
\end{eqnarray}
The associated BdG Floquet operator can then be written as
\begin{equation}
\mathcal{U} = e^{-\mathrm{i} \frac{\mathcal{H}_{(1/2),2}^{(FTSC)}T}{2}} e^{-\mathrm{i} \frac{\mathcal{H}_{(1/2),1}^{(FTSC)}T}{2}} \;. \label{bdgflo}
\end{equation}}

Under periodic boundary conditions, Eq.~(\ref{bdgflo}) can be written in momentum space as
\begin{eqnarray}
\mathcal{U} &=& \sum_k \mathcal{U}_k\otimes |k\rangle \langle k| \;, \nonumber \\
\mathcal{U}_k &=& e^{-\mathrm{i} \frac{MT}{2} \tau_y} e^{-\mathrm{i} \sum_{\ell=1,2} \left(\mu_\ell T \sigma_z \frac{1+(3-2\ell)\tau_z}{4}-(J_\ell \cos(k) \sigma_z -\Delta_\ell \sin(k) \sigma_y) T \frac{1+(3-2\ell)\tau_z}{4} \right)} \;.
\end{eqnarray}
We are particularly interested in the unitary $\mathcal{U}_k' \equiv e^{\mathrm{i} \frac{MT}{4} \tau_y} \mathcal{U}_k e^{-\mathrm{i} \frac{M T}{4} \tau_y}$, which represents the momentum space Floquet operator in the symmetric time-frame \cite{CFTI}. It can be rewritten as
\begin{eqnarray}
\mathcal{U}_k' &=& \mathcal{F}\mathcal{G} \;, \nonumber \\
\mathcal{F} &=& e^{-\mathrm{i} \frac{M T}{4} \tau_y} e^{-\mathrm{i} \sum_{\ell=1,2} \left(\mu_\ell T \sigma_z \frac{1+(3-2\ell)\tau_z}{8}-(J_\ell \cos(k) \sigma_z -\Delta_\ell \sin(k) \sigma_y) T \frac{1+(3-2\ell)\tau_z}{8} \right)} \;, \nonumber \\
\mathcal{G} &=& \mathcal{C}^\dagger \mathcal{F}^\dagger \mathcal{C} = e^{-\mathrm{i} \sum_{\ell=1,2} \left(\mu_\ell T \sigma_z \frac{1+(3-2\ell)\tau_z}{8}-(J_\ell \cos(k) \sigma_z -\Delta_\ell \sin(k) \sigma_y) T \frac{1+(3-2\ell)\tau_z}{8} \right)} e^{-\mathrm{i} \frac{M T}{4} \tau_y} \;,
\end{eqnarray}
where $\mathcal{C}=\sigma_x \tau_z$ is the chiral symmetry operator. Indeed, $\mathcal{C} \mathcal{U}_k' \mathcal{C}^\dagger =\mathcal{U}_k'^\dagger$. In the following, we focus on the ideal square-root limit $MT=\pi$. For simplicity, we take $\mu_2 = -2J_2 = -2\Delta_2 = m \mu_1= 2m J_1 = 2m\Delta_1\equiv m\delta$ with $m\in \mathbb{R}$. In the canonical basis which diagonalizes $\mathcal{C}$, we may write
\begin{equation}
\mathcal{F} = \left(\begin{array}{cc}
A & B \\
C & D
\end{array} \right)\;,
\end{equation}
where $A$ and $C$ are $2\times 2$ matrices whose exact expressions are not very relevant to our discussion, whereas
\begin{eqnarray}
B &=& \frac{1}{2} \left( \begin{array}{cc}
c_+ -\mathrm{i} s_- & c_- +\mathrm{i} s_+ \\
c_- -\mathrm{i} s_+ & c_+ +\mathrm{i} s_-
\end{array}\right) \;, \nonumber \\
D &=& \frac{1}{2} \left( \begin{array}{cc}
c_+ +\mathrm{i} s_- & c_- +\mathrm{i} s_+ \\
c_- -\mathrm{i} s_+ & c_+ -\mathrm{i} s_-
\end{array}\right) \;, \nonumber \\
c_\pm&=& \cos\left(\frac{m\delta T}{4} \sqrt{2(1+\cos k)} \right) \pm \cos\left(\frac{\delta T}{4} \sqrt{2(1-\cos k)} \right) \;, \nonumber \\
s_\pm &=& \sin\left(\frac{\delta T}{4} \sqrt{2(1-\cos k)} \right) \frac{1-e^{-\mathrm{i} k}}{\sqrt{2(1-\cos(k))}}\pm \sin\left(\frac{m\delta T}{4} \sqrt{2(1+\cos k)} \right) \frac{1+e^{\mathrm{i} k}}{\sqrt{2(1+\cos(k))}} \;.
\end{eqnarray}

By defining $z=\frac{\delta T}{4}\left(\sqrt{2(1-\cos k)} -\mathrm{i} \; m \sqrt{2(1+\cos k)} \right)$, it follows that the topological invariants $\nu_0^{(1/2)}$ and $\nu_\pi^{(1/2)}$ can be evaluated as \cite{FMF3}
\begin{eqnarray}
\nu_0^{(1/2)} &=& \frac{1}{2\pi \mathrm{i} } \int_{-\pi}^{\pi} dk \mathrm{Tr} \left( B^{-1} \frac{dB}{dk} \right) \;, \nonumber \\
&=& \frac{1}{2\pi \mathrm{i} } \int_{-\pi}^{\pi} dk \frac{1}{\cos(\mathrm{Re}(z))\cos(\mathrm{Im}(z))+\mathrm{i} \sin(\mathrm{Re}(z))\sin(\mathrm{Im}(z))}  \frac{d\left[\cos(\mathrm{Re}(z))\cos(\mathrm{Im}(z))+\mathrm{i} \sin(\mathrm{Re}(z))\sin(\mathrm{Im}(z))\right]}{dk}  \;, \nonumber \\
\nu_\pi^{(1/2)} &=& \frac{1}{2\pi \mathrm{i} } \int_{-\pi}^{\pi} dk \mathrm{Tr} \left(D^{-1} \frac{dD}{dk} \right) \;, \nonumber \\
&=& \frac{1}{2\pi \mathrm{i} } \int_{-\pi}^{\pi} dk \frac{1}{\cos(\mathrm{Re}(z))\cos(\mathrm{Im}(z))+\mathrm{i} \sin(\mathrm{Re}(z))\sin(\mathrm{Im}(z))}  \frac{d\left[\cos(\mathrm{Re}(z))\cos(\mathrm{Im}(z))+\mathrm{i} \sin(\mathrm{Re}(z))\sin(\mathrm{Im}(z))\right]}{dk} \;. \nonumber \\
\end{eqnarray}
That is, $\nu_0^{(1/2)}=\nu_\pi^{(1/2)}$.To make further progress, we note that 
\begin{equation}
d\left[\cos(\mathrm{Re}(z))\cos(\mathrm{Im}(z))+\mathrm{i} \sin(\mathrm{Re}(z))\sin(\mathrm{Im}(z))\right] = -\sin(\mathrm{Re}(z))\cos(\mathrm{Im}(z)) dz^* +\mathrm{i} \cos(\mathrm{Re}(z))\sin(\mathrm{Im}(z)) dz \;.   
\end{equation}
By further turning the momentum integration into a complex contour integration, i.e., $\int_{-\pi}^\pi \rightarrow \frac{1}{2} \oint $ \cite{RWB2}, we obtain
\begin{equation}
\nu_0^{(1/2)}=\nu_\pi^{(1/2)} = -\frac{1}{4\pi \mathrm{i}} \oint \frac{\sin(\mathrm{Re}(z))\cos(\mathrm{Im}(z)) dz^* -\mathrm{i} \cos(\mathrm{Re}(z))\sin(\mathrm{Im}(z)) dz}{\cos(\mathrm{Re}(z))\cos(\mathrm{Im}(z))+\mathrm{i} \sin(\mathrm{Re}(z))\sin(\mathrm{Im}(z))} \;.
\end{equation}
Note that this is exactly the same expression as $\nu_\pi$ of the original (squared) model, i.e., Eq.~(A2) in Ref.~\cite{RWB2}. This proves that the number of MPMs in the original model translates to the same number of MZMs and MPMs in the square-rooted model.

\RB{\section{B. Square-root continuously-driven topological superconductors} \label{appB}
To demonstrate the general applicability of our construction, we now apply our square-root procedure to some continuously-driven variations of Floquet topological superconductors. To this end, we consider a generic time-periodic Hamiltonian of the form
\begin{equation}
H(t) = \sum_{j=1}^N \mu(t) c_j^\dagger c_j +\sum_{j=1}^{N-1} \left(-J(t) c_j^\dagger c_{j+1} +\Delta(t) c_j^\dagger c_{j+1}^\dagger +h.c. \right)\;. \label{contFTSC}
\end{equation} 
For brevity, we focus on two representative examples of driving: (i) $\mu(t)= \mu(1+\cos(\omega t)), J(t)=J,\Delta(t)=\Delta$, and (ii) $\mu(t)=\mu[1+m(1+\cos(\omega t))]$, $J(t)=J[1-m(1+\cos(\omega t))],\Delta(t)=\Delta[1-m(1+\cos(\omega t))]$.} 

\begin{center}
	\begin{figure}
		\includegraphics[scale=1]{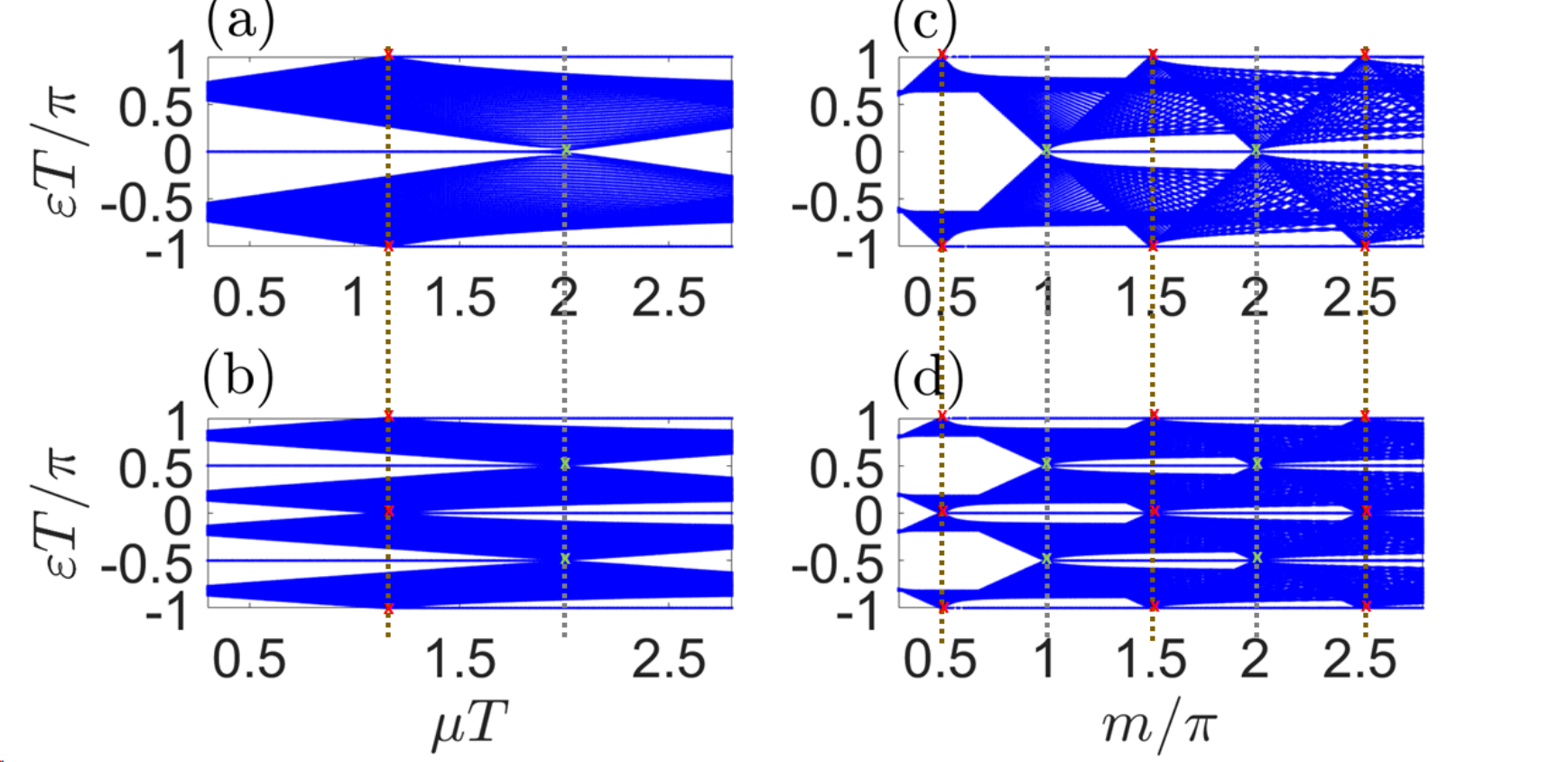}
		\caption{Quasienergy excitation spectrum of continuously-driven topological superconductors (a,c) and their square-root counterpart (b,d). Panel (a,b) corresponds to case (i), whereas (c,d) describes case (ii). System parameters are chosen as $MT=\pi$ and (a,b) $JT=\Delta T=1$, (c,d) $\mu T=2JT=2 \Delta T =1$.}
		\label{fig:app0}
	\end{figure}
\end{center}

\RB{Case (i) represents a physically relevant scenario in which chemical potential is the only parameter to be modulated due to it being the most controllable. As shown in Fig.~\ref{fig:app0}(a), both MZMs and MPMs can still emerge at some chemical potential amplitudes in such a simple driven system. To apply our square-root procedure, we take Eq.~(\ref{contFTSC}) as $H(t)$ and $H(t+\frac{T}{2})$ in Eq.~(2) of the main text. The quasienergy excitation spectrum of the obtained square-root model is presented in Fig.~\ref{fig:app0}(b). Upon side-by-side comparison between panel (a) and (b) of Fig.~\ref{fig:app0}, the same conclusion as that in the main text is obtained; the topological phase transitions at zero ($\pi/T$) quasienergy excitation in the original system translate into topological phase transitions at $\pm \pi/(2T)$ (zero and $\pi/T$).}

\RB{Case (ii) describes a continuous variation of the model presented in the main text; by periodically driving all system parameters, the imbalance between chemical potential and hopping/pairing amplitudes, i.e., the parameter $m$, can induce multiple topological phase transitions that result in arbitrarily many MZMs and MPMs (see Fig.~\ref{fig:app0}(c)). The square-root version of such a model then yields a quasienergy excitation spectrum of Fig.~\ref{fig:app0}(d), which indeed supports arbitrarily many MZMs, MPMs, and $\pi/2$ modes.}

\RB{\section{C. Detailed Analysis of square-root FTCs}}
\label{app1}

\RB{\subsection{Physical and robust implementation of square-root FTCs}}

The proposed square-rooting procedure amounts to introducing an additional spin-1/2 particle interacting with the whole $N$ particles in a 1D lattice, which may appear unrealistic in the thermodynamic limit due to the nonlocality of the interaction between two spins that are very far apart. In the following, we will show that in both MBL-protected and disorder-free FTCs considered in the main text, the resulting $N+1$ particles systems can instead be realized from two spin-1/2 chains, each of size $N$, coupled only via two-body nearest-neighbor interactions. The main idea of this mapping is through the identification of $N-1$ stabilizer (mutually commuting) operators that can be fixed to a specific value, thus bringing the $2^{2N}$-dimensional Hilbert space of the two interacting spin chains down to a dimension of $2^{N+1}$. By identifying a set of logical operators (effective Pauli matrices) acting within this $2^{N+1}$ subspace, square-root of Eqs.~(5) and (6) in the main text can be implemented in a more physical and robust way, as further demonstrated below. 

To be more specific, consider two spin-1/2 chains of size $N$ characterized by a set of Pauli matrices $X_{j,S}$ and $Z_{j,S}$ where $j=1,\cdots,N$ and $S=A,B$ label the lattice site and species respectively. Define the mutually commuting stabilizer operators $\mathcal{S}_k=Z_{k,A}Z_{k,B}Z_{k+1,A}Z_{k+1,B}$ for $k=1,\cdots,N-1$. It follows that the operators $\overline{X}_j= X_{j,A}X_{j,B}$ and $\overline{Z}_j =Z_{j,A}$, where $j=1,\cdots, N$, as well as $\overline{Y}_{N+1}=\prod_{j=1}^N X_{j,B}$ and $\overline{Z}_{N+1}=Z_{1,A}Z_{1,B}$ commute with all $\mathcal{S}_k$. Moreover, these operators fulfill the Pauli algebra $\left\lbrace \overline{X}_{j}, \overline{Z}_{j}\right\rbrace =0$ and $[\overline{X}_{j}, \overline{Z}_{j'}]=[\overline{Z}_{j}, \overline{Z}_{j'}]=[\overline{X}_{j}, \overline{X}_{j'}]=0$ for $j\neq j'$, thus spanning a $2^{N+1}$-dimensional Hilbert space.

Square-root of Eqs~(5) and (6) in the main text may now be realized with a system of two spin chains by identifying $X_j\rightarrow \overline{X}_j$, $Z_j \rightarrow \overline{Z}_j$, $\tau_z \rightarrow \overline{Z}_{N+1}$, and $\tau_y \rightarrow \overline{Y}_{N+1}$, thus resulting in the Floquet operators
\begin{eqnarray}
\tilde{u}_{(1/2)}^{(MBL)} &=& \exp\left(-\mathrm{i} \frac{MT}{2} \prod_{j=1}^N X_{j,B}\right) \exp\left(-\mathrm{i} \sum_j \left[  \left(\frac{J_j T}{4} Z_{j,A} Z_{j+1,A} + \frac{h_j^Z T}{4} Z_{j,A}\right) (1-Z_{1,A}Z_{1,B}) \right.\right. \nonumber \\
&+& \left. \left. \frac{h_j T}{4} X_{j,A}X_{j,B} (1+Z_{1,A}Z_{1,B})\right]\right) \;, \nonumber \\
\tilde{u}_{(1/2)}^{(LMG)} &=& \exp\left(-\mathrm{i} \frac{MT}{2} \prod_{j=1}^N X_{j,B}\right)\RB{\left[\mathcal{T} \exp\left(-\mathrm{i} \int_0^{T/2} \tilde{h}(t) dt\right) \right] } \;, \label{Eq:app1}
\end{eqnarray}
\RB{where
\begin{equation}
\tilde{h}(t) = \sum_j \left\lbrace \left( \sum_i \frac{J}{2N} Z_{i,A} Z_{j,A} + (h+\phi) X_{j,A}X_{j,B} \right)+\sum_i \frac{J\cos(\omega t)}{2N} Z_{i,A} Z_{j,A} Z_{1,A}Z_{1,B} +h\cos(\omega t) X_{j,A} X_{j,B} Z_{1,A}Z_{1,B}\right\rbrace \;. 
\end{equation}}
While $\tilde{u}_{(1/2)}^{(MBL)}$ and $\tilde{u}_{(1/2)}^{(LMG)}$ are $2^{2N}$-dimensional, they commute with $\mathcal{S}_k$ and can thus be written as block diagonal matrices, where each block is of dimension $2^{N+1}$. That is, $\tilde{u}_{(1/2)}^{(MBL)}$ and $\tilde{u}_{(1/2)}^{(LMG)}$ actually encode multiple copies of the Floquet operator associated with the square-root of Eqs.~(5) and (6) in the main text, a fact that can be exploited to further simplify the systems and make them more robust in the following.

First, we note that the perfect square-root limit corresponds to setting $MT=\pi$ in Eq.~(\ref{Eq:app1}), so that $\exp\left(-\mathrm{i} \frac{MT}{2}\prod_{j=1}^N X_{j,B}\right)=-\mathrm{i} \prod_{j=1}^N X_{j,B}$. Note that the latter can be equivalently written as $\exp\left(-\mathrm{i} \frac{MT}{2}\sum_{j=1}^N X_{j,B}\right)$, which is not only more realistic due to it consisting of terms involving only a single Pauli matrix rather than an $N$-body Pauli interaction, but it is also more robust against imperfection in $M$ \cite{repDTC}. Second, since $\mathcal{S}_k$ are good quantum numbers, we may introduce them anywhere in Eq.~(\ref{Eq:app1}) whenever appropriate to simplify it. In particular, this allows us to modify $Z_{1,A}Z_{1,B}\rightarrow \prod_{k=1}^{j-1} \mathcal{S}_{k} Z_{1,A}Z_{1,B} = Z_{j,A}Z_{j,B}$ in the second exponential of Eq.~(\ref{Eq:app1}). Together, these yield the improved Floquet operators
\begin{eqnarray}
u_{(1/2)}^{(MBL)} &=& \exp\left(-\mathrm{i} \frac{MT}{2} \sum_{j=1}^N X_{j,B}\right) \exp\left(-\mathrm{i} \sum_j \left[  \left(\frac{J_{j,\rm intra} T}{4} Z_{j,A} Z_{j+1,A} + \frac{J_{j,\rm inter} T}{4} Z_{j,B} Z_{j+1,A}+ \sum_{S=A,B} \frac{h_{j,S}^Z T}{4} Z_{j,S}\right) \right. \right.  \nonumber \\
&+& \left. \left. \frac{h_{X,j} T}{4} X_{j,A}X_{j,B} + \frac{h_{Y,j} T}{4} Y_{j,A}Y_{j,B}\right]\right) \;, \nonumber \\
u_{(1/2)}^{(LMG)} &=& \exp\left(-\mathrm{i} \frac{M T}{2} \sum_{j=1}^N X_{j,B}\right) \RB{\left[\mathcal{T} \exp\left(-\mathrm{i} \int_0^{T/2} h(t) dt\right) \right] } \;, \label{Eq:app2}
\end{eqnarray}
\RB{where
	\begin{equation}
	h(t) = \sum_j \left\lbrace \left( \sum_i \frac{J}{2N} Z_{i,A} Z_{j,A} + (h+\phi) X_{j,A}X_{j,B} \right)+\sum_i \frac{J\cos(\omega t)}{2N} Z_{i,A} Z_{j,B} +h\cos(\omega t) Y_{j,A} Y_{j,B} \right\rbrace \;, \label{LMGHam}
	\end{equation}}
and we have further split the system parameters $J_j$, $h_j^Z$, and $h_j$ in $u_{(1/2)}^{(MBL)}$ into $J_{j,\rm intra}$, $J_{j,\rm inter}$, $h_{j,A}^Z$, $h_{j,B}^Z$, $h_{X,j}$, and $h_{Y,j}$ to allow imperfection from realizing the terms $(1+\tau_z)$ and $(1-\tau_z)$ in Eq.~(2) of the main text. 


\begin{center}
	\begin{figure}
		\includegraphics[scale=1]{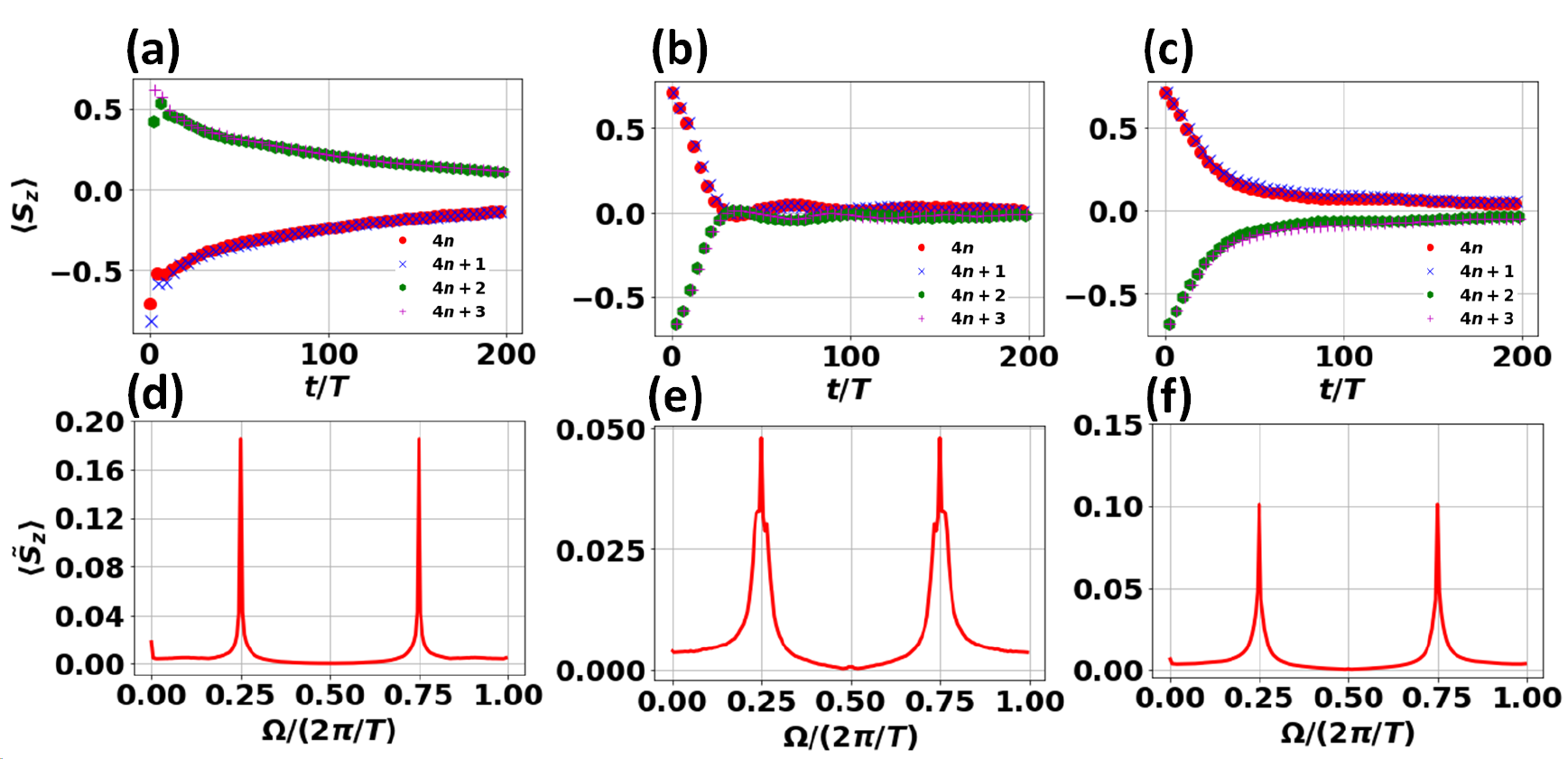}
		\caption{(a) Stroboscopic magnetization profile under $u_{(1/2)}^{(MBL)}$ for a system of two interacting spin-1/2 chains, each of size $4$ and starting with $|\psi(0)\rangle = \prod_{j=1}^N e^{-\mathrm{i} \frac{\pi}{8} Y_j} |00\cdots 0\rangle$. (b,c) Stroboscopic magnetization profile under the square-root of Eq.~(5) in the main text for a spin chain of size (b) $4$ and (c) $8$ particles. Panels (d,e,f) show the corresponding power spectrum. Each data point is averaged over $500$ disorder realizations. The system parameters are chosen as $\bar{h}_X T=\bar{h}_Y T =0.95 \pi$, $\bar{J}_{\rm intra} T=\bar{J}_{\rm inter} T=2$, $\bar{h}_A^Z T = \bar{h}_B^Z T = 0.3$, $\bar{M} T =0.95\pi$, $\Delta h_X T=\Delta h_Y T=0.05\pi$, $\Delta J_{\rm intra} T = \Delta J_{\rm inter} T = 1$, $\Delta h_A^Z T = \Delta h_B^Z T = 0.3$, and $\Delta M T=0.05\pi$.}
		\label{fig:app1}
	\end{figure}
\end{center}

\begin{center}
	\begin{figure}
		\includegraphics[scale=1]{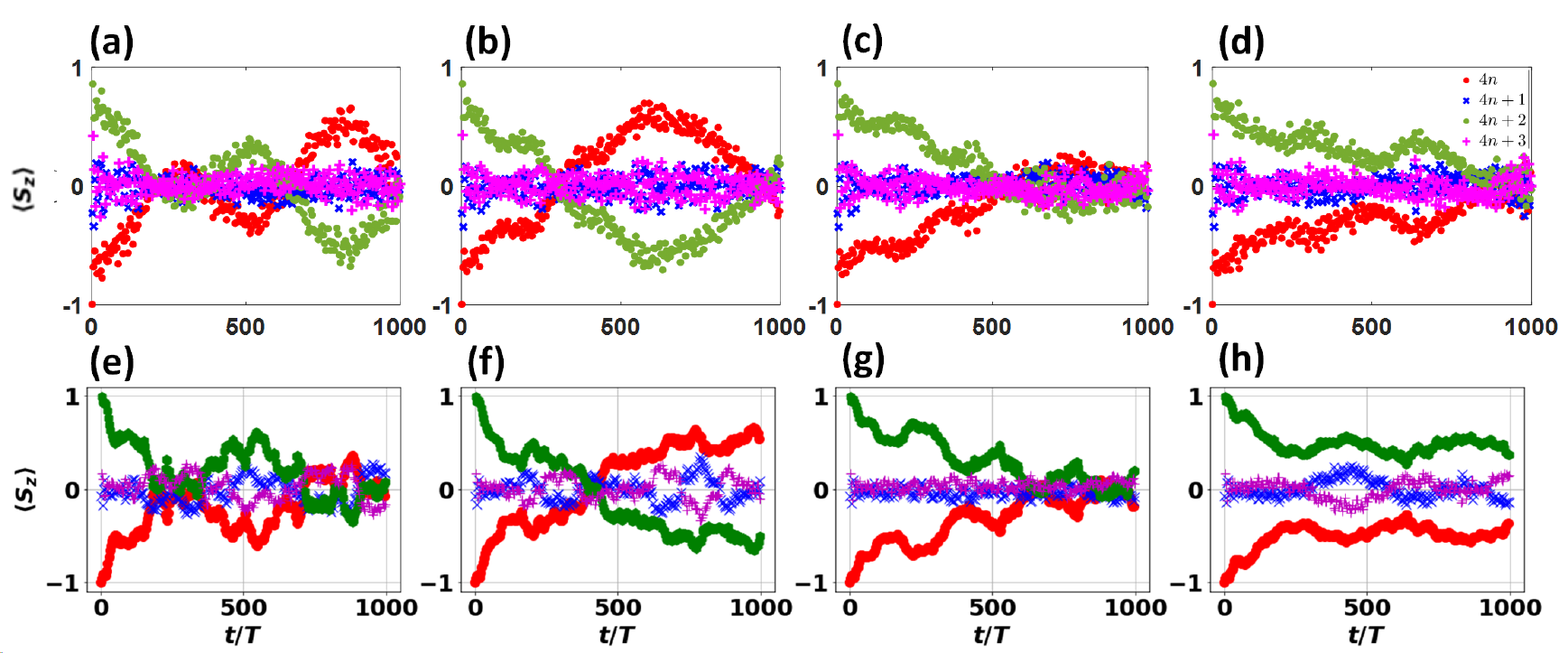}
		\caption{\RB{Stroboscopic magnetization profile under (a-d) the square-root of Eq.~(6) in the main text and (e-h) $u_{(1/2)}^{(LMG)}$ for a system of (a,e) $8$, (b,f) $10$, (c,g) $12$, and (d,h) $14$ particles. The system parameters are the same as those in Fig. 3 of the main text.}}
		\label{fig:app2}
	\end{figure}
\end{center}

To demonstrate the better robustness of the two interacting chains models derived here as compared with the direct square-root of Eq.~(5) and Eq.~(6) in the main text, the magnetization profile for both versions of each model at the same parameter values is presented in Fig.~\ref{fig:app1} and Fig.~\ref{fig:app2}. For a comprehensive comparison, two different system sizes are considered for the case of the square-root MBL-protected FTC. Specifically, the system size of Fig.~\ref{fig:app1}(b) is chosen as it leads to the same Hilbert space dimension as that of $u_{(1/2)}^{(MBL)}$ in panel (a) when restricted to a stabilizer subspace of fixed $\mathcal{S}_k$. On the other hand, the system size of Fig.~\ref{fig:app1}(c) is chosen so that the resulting spin chain contains the same number of particles, i.e., $8$, as that of $u_{(1/2)}^{(MBL)}$ in panel (a). Notably, while Fig.~\ref{fig:app1}(c) demonstrates a stronger $4T$ oscillation over Fig.~\ref{fig:app1}(b) due to its larger system size, Fig.~\ref{fig:app1}(a) outperforms both Figs.~\ref{fig:app1}(b) and (c).

\RB{Since $u_{(1/2)}^{(LMG)}$ no longer conserves the total spin number, very large system sizes are no longer accessible via exact diagonalization. Therefore, in Fig.~\ref{fig:app2}, we provide a side-by-side comparison between the direct square-root of Eq.~(6) in the main text and $u_{(1/2)}^{(LMG)}$ at several relatively small system sizes. Remarkably, a clear and long-lived $4T$-oscillation pattern can already be observed in the improved model with as small as $14$ particles.} This in turn demonstrates the potential of our square-rooting procedure to observe disorder-free large period FTCs at small system sizes, as claimed in the main text.    

\RB{\subsection{Origin of time-crystallinity}}

\RB{We will now discuss the main mechanism that enables our square-root FTCs to display $4T$ periodicity at a much smaller system size than that in the system of Ref.~\cite{Pizzi}. To this end, we first note that the Floquet eigenstates of a period-$n$-tupling FTC take the form of macroscopic cat states. Indeed, suppose that the Floquet operator of such a system, i.e., $u$, possesses the following set of eigenstates,
\begin{equation}
|\psi_m \rangle = \frac{1}{\sqrt{n}} \sum_{s=1}^n \exp\left(-\mathrm{i} \frac{2\pi s m}{n}\right) |ss\cdots s\rangle \;, \label{cat}
\end{equation}
where $m=0,1,\cdots, n-1$, $|\psi_m \rangle$ and $|\psi_{m+1} \rangle$ are separated in quasienergy by $\Delta \varepsilon= \frac{2\pi}{nT}$. Over one period, we then have 
\begin{equation}
|ss\cdots s\rangle = \frac{1}{\sqrt{n}} \sum_{m=0}^{n-1} |\psi_m \rangle \rightarrow \frac{1}{\sqrt{n}} \sum_{m=0}^{n-1} \exp\left(-\mathrm{i} \left(\varepsilon_0 T+ \frac{2\pi m}{n} \right) \right) |\psi_m \rangle \neq |ss\cdots s\rangle \;,
\end{equation}
where $\varepsilon_0$ is the quasienergy associated with $|\psi_0\rangle$, i.e., $u|\psi_0\rangle =e^{-\mathrm{i} \varepsilon_0 T}|\psi_0\rangle$. That is, an initial state $|ss\cdots s\rangle$ transforms into a different state due to the relative phase difference between its Floquet eigenstate constituents. Due to the same mechanism, the state further evolves into various other distinct states after $2,3,\cdots,$ and $n-1$ periods. Over $n$ periods, however, this relative phase difference will have accumulated into an integer multiple $2\pi$. In this case, the state returns to $|ss\cdots s\rangle$, establishing the $nT$-periodicity. If all, or at least a large number, of the system's Floquet eigenstates take the form of Eq.~(\ref{cat}), the above argument applies to any generic physical inital state.}

\RB{Since Eq.~(\ref{cat}) comprises a superposition of tensor products of states with $n$ degrees of freedom, it only naturally arises in a chain of $n$-level particles. Therefore, existing experiments, which only have access to $2$-level and $3$-level particles, are natively limited to support period-doubling \cite{DTCexp1,DTCexp7,DTCexp8} and period-tripling \cite{DTCexp2} FTCs. While our square-root FTCs also consist of $2$-level particles, the arrangement of the systems into two 1D lattices and the presence of intersite interactions beyond the usual $ZZ$-type allow for the grouping of two $2$-level particles into an effective $4$-level particle, which then leads to a native $4T$-periodicity by the argument above.}

\RB{More quantitatively, we can analytically verify the formation of eigenstates of the form Eq.~(\ref{cat}) in our square-root FTCs at special parameter values, i.e., $u_{(1/2)}^{(MBL)}$ at $J_{j,\rm intra}=J_{j,\rm inter}=h_{j,s}^Z=0$, $MT=h_{X,j}T=h_{Y,j}T =\pi$ or $u_{(1/2)}^{(LMG)}$ at $J=h=0$, $\phi T= \pi/2$. For brevity, we will only focus on the square-root FTC described by $u_{(1/2)}^{(MBL)}$ in the following as the analysis for $u_{(1/2)}^{(LMG)}$ proceeds in the same way. To this end, let $|s_{1,A}s_{1,B}s_{2,A}s_{2,B}\cdots s_{N,A}s_{N,B}\rangle$ denote a general basis state such that the particle at the $j$th site of chain $A$ ($B$) is in the state $s_{j,A}$ ($s_{j,B}$), where $s_{j,A},s_{j,B}\in\left\lbrace 0,1 \right\rbrace$. It is instructive to define a shorthand notation $|\mathbf{0}_j\rangle = |0_{j,A}0_{j,B}\rangle $, $|\mathbf{1}_j\rangle = |0_{j,A}1_{j,B}\rangle $, $|\mathbf{2}_j\rangle = |1_{j,A}1_{j,B}\rangle $, and $|\mathbf{3}_j\rangle = |1_{j,A}0_{j,B}\rangle $, so that the general basis state above can be written as $|\mathbf{s}_1 \mathbf{s}_2\cdots \mathbf{s}_N \rangle$ with $\mathbf{s}_j\in\left\lbrace 0,1,2,3 \right\rbrace$. At the special parameter values under consideration, it is easily verified that  
\begin{equation}
u_{(1/2)}^{(MBL)} |\mathbf{s}_1 \mathbf{s}_2\cdots \mathbf{s}_N \rangle = (-1)^N \left( \prod_{j=1}^N \mathrm{i} \frac{1+(-1)^{\mathbf{s}_j}}{2}\right)  |(\mathbf{s}_1 \oplus \mathbf{1}) (\mathbf{s}_2 \oplus \mathbf{1})\cdots (\mathbf{s}_N \oplus \mathbf{1}) \rangle \;, \label{action}
\end{equation}
where $\oplus$ is the addition modulo $4$. Let $L$ be the number of $\mathbf{s}_1,\cdots,\mathbf{s}_N$ that take even numbers, i.e., $0$ or $2$. The prefactor in the right hand side of Eq.~(\ref{action}) then reduces to $(-1)^N \mathrm{i}^L$. From Eq.~(\ref{action}), the following eigenstates can be identified, 
\begin{equation}
|\psi_{\mathbf{s},m} \rangle = \frac{1}{2} \left(|\mathbf{s}\rangle + e^{-\mathrm{i} \frac{\pi}{4}\left(N-2L-2m\right)} |\mathbf{s} \oplus \mathbf{1} \rangle +(-1)^m|\mathbf{s} \oplus \mathbf{2}\rangle +e^{-\mathrm{i} \frac{\pi}{4}\left(N-2L+2m\right)} |\mathbf{s} \oplus \mathbf{3}\rangle \right) \;,
\end{equation}
where $m=0,1,2,3$, the associated quasienergies are $\varepsilon_m=\frac{\pi\left(3N-4+2m\right)}{4T}$, and we have defined $|\mathbf{s} \oplus \mathbf{k} \rangle = |(\mathbf{s}_1 \oplus \mathbf{k})(\mathbf{s}_2 \oplus \mathbf{k})\cdots (\mathbf{s}_N \oplus \mathbf{k}) \rangle$. Since $|\psi_{\mathbf{s},m} \rangle$ can be constructed out of any basis state $|\mathbf{s}\rangle$, we conclude that \emph{all} eigenstates of $u_{(1/2)}^{(MBL)}$ indeed take the macroscopic cat states with $n=4$, as expected from a period-quadrupling time crystal. By turning on the interaction $J_{j,\rm intra},J_{j,\rm inter}$, the above cat state structure will then exhibit robustness against perturbation in the system parameters $h_{j,s}^Z$, $M$, $h_{X,j}$, and $h_{Y,j}$ as numerically demonstrated above.}

\RB{Finally, we expect that a similar mechanism may be at play in the system of Ref.~\cite{Pizzi} to enable the emergence of large period FTCs. Indeed, the long range interaction considered in such a system may naturally facilitate the grouping of two-level particles into a larger-level particle. However, due to the lack of additional interactions beyond the $ZZ$-type, it may take on average more than two two-level particles to form an effective four-level particle, thus resulting in a much larger system size requirement for observing period-quadrupling FTCs in Ref.~\cite{Pizzi}.} 
	

\RB{\subsection{Potential experimental realization of square-root FTCs}}

It is noted that all terms appearing in Eq.~(\ref{Eq:app2}) can be readily implemented in trapped ions \cite{DTCexp1,DTCexp7} or superconducting circuit \cite{DTCexp8} setups. \RB{Indeed, the long range nature of the driven LMG system is naturally achieved in trapped ions experiments \cite{DTCexp1,DTCexp7}. There, long range $ZZ$ interaction is realized via the spin-dependent optical dipole forces. Other types of interactions that appear in Eq.~(\ref{LMGHam}), i.e., $XX$ and $YY$ interactions, can then be obtained by applying appropriate $\pi/4$ pulses, realized via optically
	driven Raman transitions between two hyperfine clock
	states of $^{171}Yb+$ ion, before and after the application of the native $ZZ$ interaction. This construction is easily understood by noting the identity 
	\begin{equation}
	e^{-\mathrm{i} \pi/4 P_1} P_2 e^{\mathrm{i} \pi/4 P_1} = \mathrm{i} P_2 P_1 \;,\label{identity}
	\end{equation}
	where $P_1$ and $P_2$ are anticommuting Pauli matrices. In the case of $P_2=Z\otimes Z$, applying Eq.~(\ref{identity}) under $P_1^{(1)}=X\otimes I$ ($P_1^{(1)}=Y\otimes I$), followed by another application of Eq.~(\ref{identity}) under $P_1^{(2)}=I\otimes X$ ($P_1^{(2)}=I\otimes Y$), yields an effective interaction of the form $Y\otimes Y$ ($X\otimes X$).} 

\RB{The MBL-protected FTC described by $u_{(1/2)}^{(MBL)}$ involves only nearest-neighbor interactions and can be more easily realized with superconducting qubits \cite{DTCexp8}. For example, Google's Sycamore processor \cite{Syc} allows the application of any single qubit rotation and the $iSWAP=e^{-\mathrm{i} \frac{\pi}{4} (X\otimes X +Y\otimes Y)}$ gate natively. By writing $u_{(1/2)}^{(MBL)}$ as a finite-depth circuit involving solely these native gates, it can then be directly implemented in such a setup. Indeed, this approach has sucessfully been carried out in an actual experiment to realize a $2T$-period MBL-protected FTC \cite{DTCexp8}, i.e., the parent model of $u_{(1/2)}^{(MBL)}$. While both $u_{(1/2)}^{(MBL)}$ and its parent model involve $ZZ$ interaction, $u_{(1/2)}^{(MBL)}$ additionally requires the presence of nearest-neighbor interactions of the form $X\otimes X$ and $Y\otimes Y$. By using Eq.~(\ref{identity}) and following the argument of the previous paragraph, such interactions can be broken down into $ZZ$ interaction and single qubit rotations. As a result, $u_{(1/2)}^{(MBL)}$ requires precisely the same ingredients as those of its parent model and can thus be implemented in the experiment of Ref.~\cite{DTCexp8} with minimal modifications.}

\RB{\section{D. Nontrivial square-root of $H(t)=H_0(1+\sin(\omega t))$}}

\RB{By directly plugging $H(t)=H_0(1+\sin(\omega t))$ into Eq.~(2) of the main text, we obtain the Floquet operator
\begin{eqnarray}
u &=& \exp\left(-\mathrm{i} \frac{\pi}{2} \tau_y \right) \cdot \left(\begin{array}{cc}
\exp\left(-\mathrm{i} \int_0^{T/2} H_0(1+\sin(\omega t)) dt\right) & \mathbf{0} \\
\mathbf{0} & \exp\left(-\mathrm{i} \int_{T/2}^{T} H_0(1+\sin(\omega t)) dt\right)
\end{array} \right)  \;, \nonumber \\
&=& \exp\left(-\mathrm{i} \frac{\pi}{2} \tau_y \right) \cdot  \exp\left(-\mathrm{i} \frac{H_0 T}{2} \left(1+\frac{2}{\pi}\tau_z \right) \right) =\exp\left(-\mathrm{i} h_{\rm eff} T\right)\;,
\end{eqnarray}
where $h_{\rm eff}$ is the effective Hamiltonian of the square-root model. By applying the Baker-Campbell-Hausdorff formula, i.e., 
\begin{equation}
e^{A} \cdot e^{B} = \exp\left(A+B + \frac{1}{2} [A,B] + \frac{1}{12} [A,[A,B]]-\frac{1}{12} [B,[A,B]] +\cdots \right), \label{BCH}
\end{equation}
we obtain
\begin{equation}
h_{\rm eff}= \frac{\omega}{4} \tau_y +\frac{H_0}{2} \left(1+\left(\frac{2}{\pi}+\frac{\pi}{6}\right)\tau_z+2\tau_x \right)+\frac{1}{3\omega} H_0^2 \tau_y +\cdots \;. \label{effh} 
\end{equation}
where $(\cdots)$ contains higher powers of $\frac{H_0}{\omega}$. In particular, even if $H_0$ consists of at most nearest-neighbor interactions, $H_0^N$ can generate up to range-$N$ interactions. Since $h_{\rm eff}$ comprises all powers of $H_0$, it is highly nonlocal and unrealistic in static systems. This shows that the square-root of $H(t)=H_0(1+\sin(\omega t))$ obtained by our procedure has no static counterpart. Intuitively, the nontriviality of $h_{\rm eff}$ is made possible by the extra degree of freedom employed in our construction. It provides a source of noncommutativity that leads to nonzero nested commutators in Eq.~(\ref{BCH}). Interestingly, upon squaring $u$, the various terms of high-power in $\frac{H_0}{\omega}$ appearing in Eq.~(\ref{effh}) precisely cancel due to a mechanism reminiscent of the dynamical decoupling. Indeed, 
\begin{eqnarray}
u^2 &=& \exp\left(-\mathrm{i} \frac{\pi}{2} \tau_y \right) \cdot  \exp\left(-\mathrm{i} \frac{H_0 T}{2} \left(1+\frac{2}{\pi}\tau_z \right) \right) \cdot \exp\left(-\mathrm{i} \frac{\pi}{2} \tau_y \right) \cdot  \exp\left(-\mathrm{i} \frac{H_0 T}{2} \left(1+\frac{2}{\pi}\tau_z \right) \right) \;, \nonumber \\
&=& -\exp\left(-\mathrm{i} \frac{H_0 T}{2} \left(1-\frac{2}{\pi}\tau_z \right) \right) \cdot  \exp\left(-\mathrm{i} \frac{H_0 T}{2} \left(1+\frac{2}{\pi}\tau_z \right) \right) = - \exp\left(-\mathrm{i} H_0 T\right) = \exp\left(-\mathrm{i} (H_0 +\pi/T) T\right) \;,
\end{eqnarray}  
thus recovering the trivial effective Hamiltonian $H_0$ (up to a $\pi/T$ energy shift) of the squared model, as it should be.}

\section{E. $4$th- and $8$th-root models}
\label{app2}

To highlight the scalability of the square-rooting procedure described in the main text, we will now explicitly repeat the procedure two more times with respect to the square-root models obtained in the main text to yield the $4$th- and $8$th-root version of their parent Floquet phases. In the following, we will focus on the square-root Floquet topological superconductor and the square-root \RB{continuously-driven} LMG time crystal in the main text as a representative example of noninteracting and interacting system respectively.

\begin{center}
	\begin{figure}
		\includegraphics[scale=1]{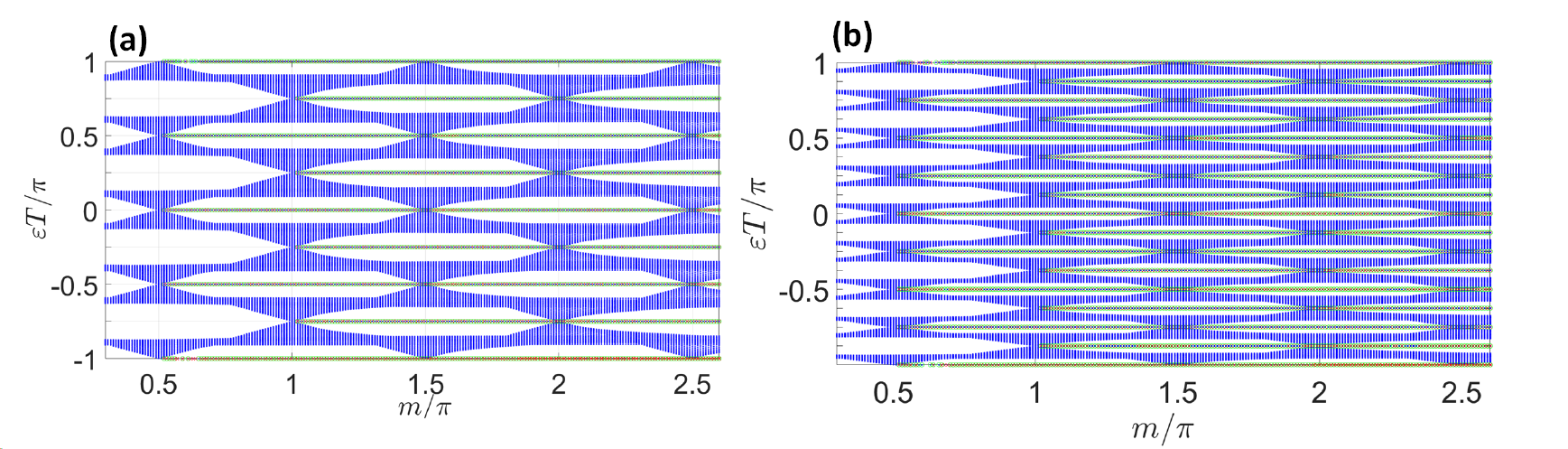}
		\caption{Quasienergy excitation spectrum associated with (a) Eq.~(\ref{4fti}) and (b) Eq.~(\ref{8fti}). The system parameters are chosen as $\mu_2 T=-2J_2 T=-2\Delta_2 T=2m\mu_1 T=2mJ_1 T= 2 m\Delta_1 T=2m$, $M T=M' T=M'' T=\pi$, and $N=50$.} 
		\label{fig:app3}
	\end{figure}
\end{center}

We start by further square-rooting the square-root Floquet topological superconductor in the main text. To this end, we introduce a new set of Pauli matrices $\tau_s'$ ($s=x,y,z$) and define a two-time-step BdG Hamiltonian $\mathcal{H}_{(1/4)}(t)$ which switches between
\begin{eqnarray}
\mathcal{H}_{(1/4),1} &=& \sum_{\ell=1,2} \sum_j \left( \mu_\ell \sigma_z \frac{1+(3-2\ell) \tau_z}{2} |j\rangle \langle j |- \left[(J_\ell \sigma_z -\mathrm{i}\Delta \sigma_y) \frac{1+(3-2\ell) \tau_z}{2} |j\rangle \langle j+1\rangle +h.c.\right] \right) \frac{1+\tau_z'}{2} \nonumber \\  
&& +\sum_j M \tau_y \frac{1-\tau_z'}{2} \;, \;\;\;\; \mathcal{H}_{(1/4),2} = \sum_j M' \tau_y' \label{4fti}
\end{eqnarray}
after every $T/2$ time interval. We then repeat the square-rooting procedure again to obtain the time-periodic BdG Hamiltonian $\mathcal{H}_{(1/8)}(t)$ which switches between
\begin{eqnarray}
\mathcal{H}_{(1/8),1} &=& \mathcal{H}_{(1/4),1} \frac{1+\tau_z''}{2} + \mathcal{H}_{(1/4),2} \frac{1-\tau_z''}{2} \;, \;\;\;\; \mathcal{H}_{(1/8),2} = \sum_j M'' \tau_y''  \label{8fti}
\end{eqnarray}
after every $T/2$ time interval, where $\tau_s''$ ($s=x,y,z$) is another set of Pauli matrices. In Fig.~\ref{fig:app3}, the quasienergy excitation spectrum of the two models reveals that the $4$th- and $8$th-root Floquet topological superconductors support $\pi/4$ and $\pi/8$ modes respectively. Moreover, in both models, the topological phase transitions occur at the same parameter values as the parent model of Ref.~\cite{RWB2}. These confirm our expectation that repeated applications of our square-rooting procedure indeed yield Floquet systems with various exotic properties, i.e., $\pi/n$ modes, the topological origin of which can be understood from the original model. 

\begin{center}
	\begin{figure}
		\includegraphics[scale=1]{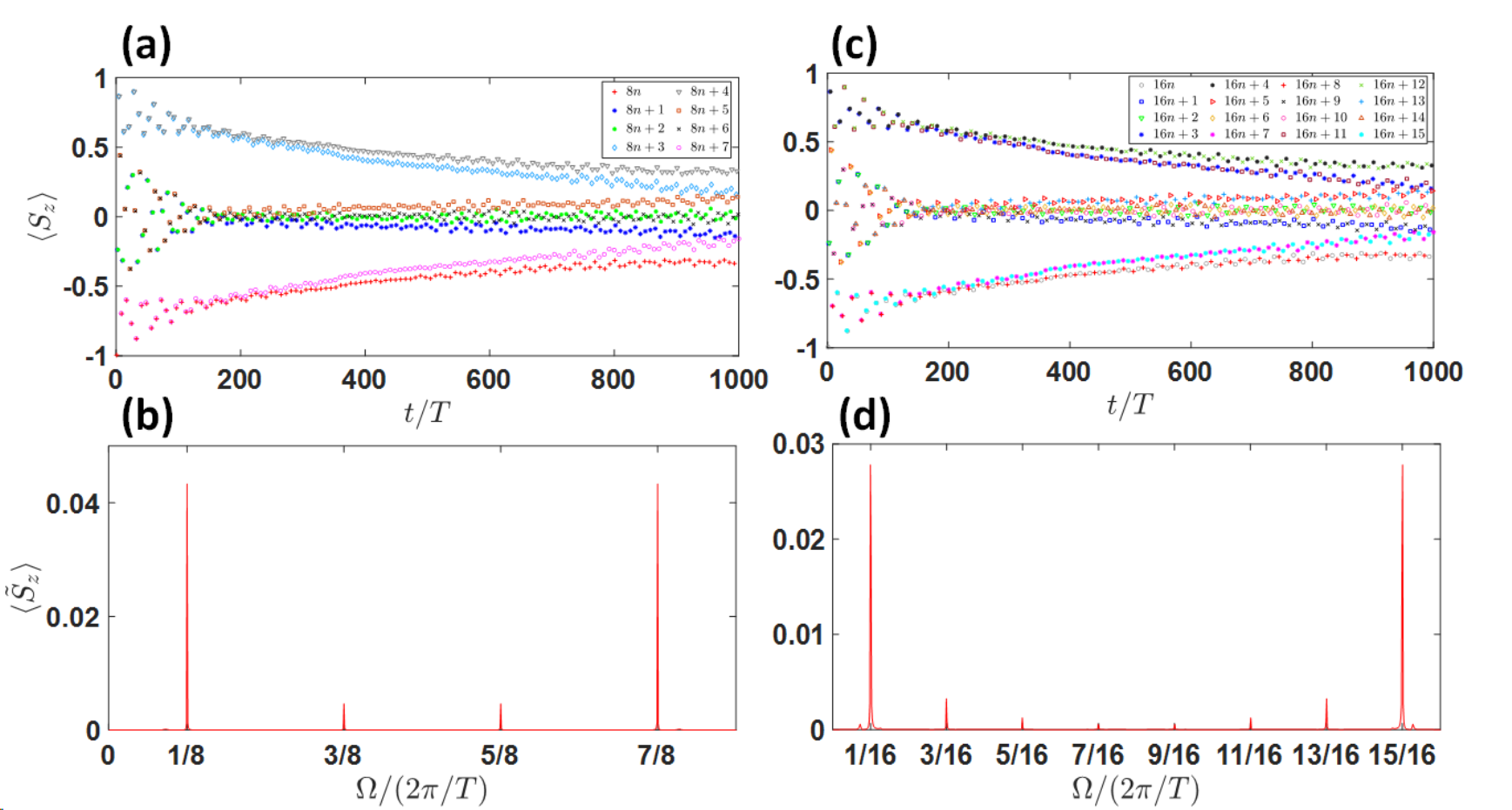}
		\caption{\RB{(a,c) Stroboscopic magnetization profile of $200$ spin-1/2 particles, starting with $|\psi(0)\rangle =|00\cdots 0\rangle$, under (a) $4$th- and (c) $8$th-root continuously-driven LMG Hamiltonian. (b,d) The associated power spectrum. The system parameters are chosen as $JT=1$, $hT=0.1$, $\phi T = \frac{0.9 \pi}{2}$, $MT=M'T=M''T=0.98\pi$.}} 
		\label{fig:app4}
	\end{figure}
\end{center}

We now turn our attention to the $4$th- and $8$th-root version of the continuously-driven LMG model. To this end, we first apply our square-rooting procedure to the square-root of Eq.~(6) in the main text to yield a $4$th-root continuously-driven LMG time crystal described by a two-time-step Hamiltonian $h^{(LMG)}_{(1/4)}(t)$ which switches between
\begin{equation}
h^{(LMG)}_{(1/4,1)}(t) = h^{(LMG)}_{(1/2,1)}(t) \frac{1+\tau_z'}{2} + h^{(LMG)}_{(1/2,2)} \frac{1-\tau_z'}{2}\;, \;\;\;\; h^{(LMG)}_{(1/4,2)} = M'\tau_y' 
\end{equation}
at every $\frac{T}{2}$ time interval. There, $h_{1/2,1}^{(LMG)}(t)$ and $h_{1/2,2}^{(LMG)}$ are respectively the first- and second-half of the period Hamiltonian of Eq.~(2) in the main text under Eq.~(6) in the main text as $H(t)$. Applying our square-rooting procedure one more time yields an $8$th-root continuously-driven LMG time crystal described by a two-time-step Hamiltonian $h^{(LMG)}_{(1/8)}(t)$ which switches between
\begin{equation}
h^{(LMG)}_{(1/8,1)}(t) = h^{(LMG)}_{(1/4,1)}(t) \frac{1+\tau_z''}{2} + h^{(LMG)}_{(1/4,2)} \frac{1-\tau_z''}{2}\;, \;\;\;\; h^{(LMG)}_{(1/8,2)} = M''\tau_y'' 
\end{equation}
at every $\frac{T}{2}$ time interval. As demonstrated in Fig.~\ref{fig:app4}, robust $8T$- and $16T$-periodic magnetization dynamics is observed for the $4$th- and $8$th-root continuously-driven model respectively, thus revealing their time crystal signature. While not shown in the figure, we have verified that the same oscillations are also observable for a system size as small as $20$. By more robustly implementing $h^{(LMG)}_{(1/4)}(t)$ and $h^{(LMG)}_{(1/8)}(t)$ with multiple chains of interacting spin-1/2 particles following the similar procedure described in Sec.~C, it is expected that such oscillations can also be observed at even smaller system sizes.

\end{document}